\documentclass{pasa}%

\title[Self-gravitating accretion discs]{The evolution of self-gravitating accretion discs}
\author[Rice]{Ken Rice$^1$ \thanks{email: wkmr@roe.ac.uk}\\
\affil{$^1$SUPA, Institute for Astronomy, Royal Observatory, Blackford Hill, Edinburgh, EH93HJ, UK}}%
\jid{PASA}
\doi{10.1017/pas.\the\year.xxx}
\jyear{\the\year}

\usepackage[authoryear]{natbib}
\bibpunct{(}{)}{;}{a}{}{,}
\begin{document}%
\begin{abstract}
It is quite likely that self-gravity will play an important role in the evolution of accretion discs, in particular
those around young stars, and those around supermassive black holes. We summarise, here, our current understanding
of the evolution of such discs, focussing more on discs in young stellar system, than on discs in active galactic nuclei.
We consider the conditions under which such discs may fragment to form bound objects, and when they might, instead, be expected
to settle into a quasi-steady, self-regulated state. We also discuss how this understanding may depend on the mass
of the disc relative to the mass of the central object, and how it might depend on the presence of external irradiation.  Additionally, we consider
whether or not fragmentation might be stochastic, where we might expect it to occur in an actual protostellar disc,
and if there is any evidence for fragmentation actually playing a role in the formation of planetary-mass bodies. 
Although there are still a number of outstanding issue, such as the convergence of simulations of self-gravitating
discs, whether or not there is more than one mode of fragmentation, and quite what role self-gravitating discs may 
play in the planet formation process, our general understanding of these systems seems quite robust.
\end{abstract}
\begin{keywords}
accretion, accretion discs -- instabilities -- planets and satellites: formation -- protoplanetary discs -- staris: formation
\end{keywords}
\maketitle%
\section{INTRODUCTION }
\label{sec:intro}
It is likely that self-gravity will, at times, play an important role in the evolution of accretion discs.  In particular
we expect it to be important in discs around active galactic nuclei \citep{nayakshin07} and in discs around very young
protostars \citep{lin87}.  Although we will discuss some general properties of self-gravitating discs, we will
focus more on protostellar discs, than on other types of potentially self-gravitating accretion discs.  

An accretion disc will be self-gravitating, or susceptible to the growth of the gravitational instability, if the 
Toomre parameter, $Q$, is close to unity \citep{safronov60,toomre64}
\begin{equation}
Q = \frac{c_s \Omega}{\pi G \Sigma} \sim 1,
\label{eq:Toomre}
\end{equation}
where $c_s$ is the sound speed, $\Omega$ is the epicyclic frequency (equal to the orbital angular frequency in Keplerian discs),
$\Sigma$ is the disc's surface density, and $G$ is the gravitational constant.  This parameter, however, only tells us
if the disc's self-gravity is important and if it will likely develop a gravitational instability.  It doesn't, however,
tell is how such discs will actually evolve. 

There are two likely pathways for a self-gravitating disc.  It could settle into a quasi-steady, self-regulated state \citep{paczynski78},
or it could become so unstable that it fragments into bound objects \citep{boss98}. The latter could be a potential star
formation process in discs around active galactic nuclei \citep{levin03}, or a mechanism for forming planets, or brown dwarfs, in discs
around young stars \citep{boss00}.  Even if a disc does not fragment, a self-gravitating phase could play a very important
role in driving angular momentum transport and, hence, allowing mass to accrete onto the central protostar \citep{lin87,rice10}.

In this paper we will summarise our current understanding of the evolution of self-gravitating accretion discs. There are a number of 
factors that likely determine the evolution of such discs.  In particular, the balance between heating and cooling, and the 
mass of the disc relative to the mass of the central object.  We'll discuss if we can establish the conditions which determine whether a 
self-gravitating disc will settle into a quasi-steady state, or fragment, and - if so - when such conditions are valid, and when they
might be violated. We'll look at the role of external factors, such as external irradiation and perturbations from
stellar, or sub-stellar, companions. We'll also discuss, in the context of protostellar discs, whether or not we expect fragmentation
to actually occur and - if so - where it is likely to occur and what we might expect the outcome to be.

The paper is structured in the following way.  In Section 2 we focus on the quasi-steady evolution of self-gravitating
discs, how such discs evolve under the influence of external irradiation, what happens if the disc mass is high relative
to the mass of the central object, and what might happen if we consider more realistic scenarios.  In Section 3 we focus
on the fragmentation of self-gravitating discs. In particular we consider what conditions determine if a disc will fragment,
when such conditions may apply and when they may be violated, could it be stochastic rather than precisely determined, and  
could fragmentation be triggered by external perturbations. In Section 4 we then discuss these results and draw conclusions.
 
\section{QUASI-STEADY EVOLUTION}
As already mentioned, a self-gravitating disc is one that has sufficient mass, or that is cold enough, 
so that the Toomre parameter, $Q$, is of order unity. Such a self-gravitating disc will generate an instability
that will dissipate and heat the disc.  If the cooling rate can match the rate at which the instability
is heating the disc, the system may then sustain a state of marginal stability \citep{paczynski78}. If not,
the system may fragment to form bound objects, potentially protoplanets or brown dwarfs in discs around young stars \cite{boss00},
or stars in discs around active galactic nuclei \citep{levin03, bonnell08}.

If such a system does maintain a state of marginal stability, one would expect the instability to transport 
angular momentum.  A standard way to describe angular momentum transport in discs is to assume that it is 
viscous and can be described using the $\alpha$-prescription \citep{shakura73}. Energy transport in a self-gravitating
disc, however, is not diffusive and so doesn't, strictly speaking, satisfy this condition \citep{balbus99}. 
Numerical studies \citep{laughlin96,nelson00,pickett00} have, however, shown that the $\alpha$-prescription can be a reasonable
way to parametrise angular momentum transport in self-gravitating discs.

\begin{figure}
\begin{center}
\includegraphics[width=8cm]{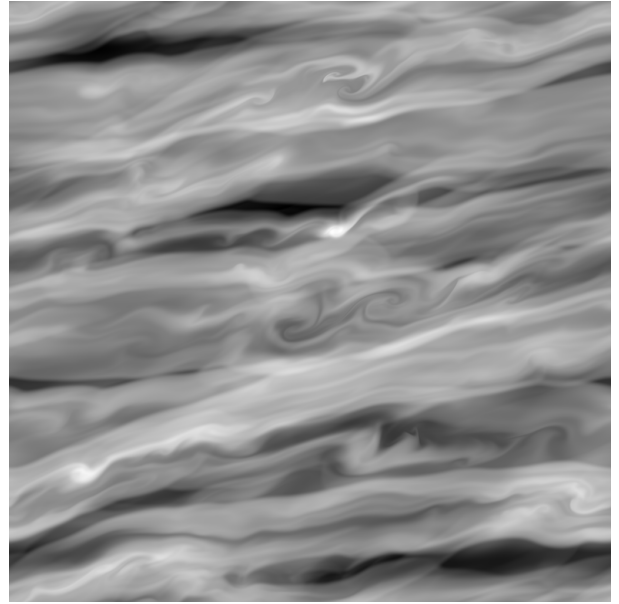}
\caption{A snapshot showing the surface density structure in a shearing sheet simulation with $\beta = 10$ and $\alpha = 2$. The disc 
has settled into a quasi-steady. self-regulated state in which heating via dissipation of the gravitational instability is balancing
the imposed cooling.}
\label{fig:slice_10_1k}
\end{center}
\end{figure}

\begin{figure}
\begin{center}
\includegraphics[width=8.5cm]{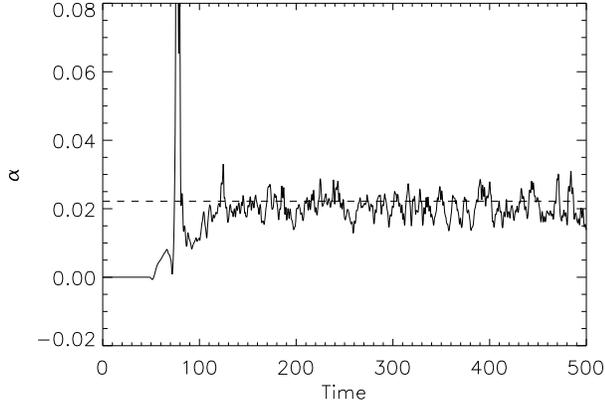}
\caption{The effective $\alpha$ plotted against time for the simulation shown in Figure \ref{fig:slice_10_1k}. Cooling is
turned on at $t = 50$, after which there is a sudden burst.  However, by $t = 100$ the simulation has settled into a quasi-steady
state with a roughly constant $\alpha$, consistent with what is expected from energy balance (dashed line).}
\label{fig:alpha_tc10}
\end{center}
\end{figure}

One of the first studies to quantify this \citep{gammie01} used a local shearing sheet in which the cooling time, $\tau$, was 
assumed to be a constant, $\beta$, divided by the local angular frequency, $\Omega$:
\begin{equation}
\tau = \beta \Omega^{-1}.
\label{eq:tau}
\end{equation}
 Such a cooling formalism has 
now become known as $\beta$-cooling.  What this work showed was that, as expected, if the disc settled into a quasi-steady state, 
the dissipation rate would match the imposed cooling, and the effective viscous $\alpha$ would then be given by
\begin{equation}
\alpha = \frac{4}{9 \gamma (\gamma - 1) \beta},
\label{eq:alpha_gammie}
\end{equation}
where $\gamma$ is the specific heat ratio. Figure \ref{fig:slice_10_1k} shows an example of the quasi-steady, surface density structure 
in a local shearing sheet simulation in which the imposed cooling is assumed to have $\beta = 10$ and in which $\gamma = 2$. Figure \ref{fig:alpha_tc10} shows 
the resulting $\alpha$ evolution. The simulation is initially run with no cooling, which is then turned on at $t = 50$. There 
is an initial burst and then the system settles into a quasi-steady state with an approximately constant $\alpha$. The 
dashed line shows the $\alpha$ value to which we'd expect it to settle, based on Equation (\ref{eq:alpha_gammie}).

Other studies employing different types of numerical methods, such as Smoothed Particle Hydrodynamics \citep{lodato04, rice05} 
and grid-based methods \citep{mejia05}, 
have similarly shown that a self-gravitating disc will tend to settle to a quasi-steady state in which $Q$ is close to 
unity and in which angular momentum transport can be described using an $\alpha$-formalism in which $\alpha$ is given by
Equation (\ref{eq:alpha_gammie}).  

There are, however, some caveats to the above that we will discuss in more detail later.  It applies in situations where the disc mass 
is low enough that the local approximation is valid, and also requires that the cooling time is not so short that the 
instability becomes non-linear and the disc breaks up into fragments. 

\subsection{Irradiation}
A fundamental assumption in some of the early work on the evolution of self-gravitating discs (e.g., \citealt{gammie01}) 
was that there was only a single heating source; dissipation of the gravitational instabilty. In such a scenario, the cooling balances
only this heating source and the effective $\alpha$ is given simply by Equation (\ref{eq:alpha_gammie}). In the presence of
an additional heating source (such as external irradiation) it is more complex, since - in a state of marginal stability - 
the cooling balances both heating due to the instability and the additional heating source.  

\begin{figure}
\begin{center}
\includegraphics[width=8.5cm]{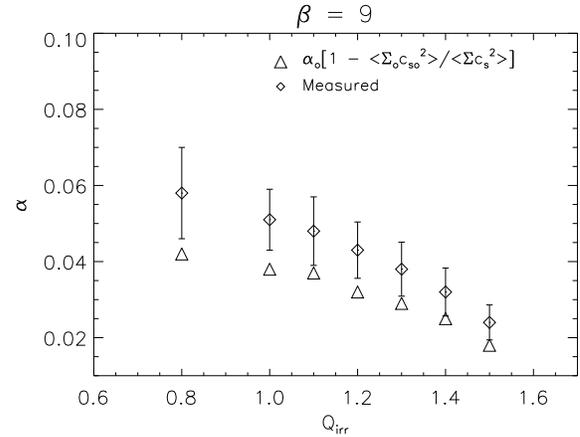}
\caption{Figure showing how the quasi-steady $\alpha$ value varies with the level external irradiation (taken from \citealt{rice11}). The triangles
are an initial semi-analytic estimate, while the diamonds with error bars are from the shearing sheet simulations. Although
there is some discrepancy, it's clear that the effective $\alpha$ decreases with increasing levels of external irradiation.}
\label{fig:alpha_Qirr}
\end{center}
\end{figure}

For a given cooling time, the effective $\alpha$ is reduced in the presence of an additional heating source, compared
to a situation where the only heating source is the gravitational instability itself. It can be shown 
\cite{rice11} that, in the presence of external irradiation, one can approximate the effective $\alpha$ using
\begin{equation}
\alpha = \frac{4}{9 \gamma (\gamma - 1) \beta} \left(1 - \frac{\langle \Sigma \rangle c^2_{so}}{\langle \Sigma c_s^2 \rangle} \right),
\label{eq:alpha_irr}
\end{equation}
where $\Sigma$ is the local surface density, $c_{so}$ is the sound speed to which the 
disc would settle in the presence of external irradiation only, and $c_s$ is the sound speed to which
the disc settles if it does tend to a state of marginal stability.  This is similar to the form
suggested by \citet{kratter10}. This can then be recast as
\begin{equation}
\alpha = \frac{4}{9 \gamma (\gamma - 1) \beta} \left(1 - \frac{Q^2_{\rm irr}}{Q^2_{\rm sat}} \right),
\label{eq:alpha_irr_Q}
\end{equation}
where $Q_{\rm irr}$ is the $Q$ value to which it would settle in the presence of external irradiation only, while $Q_{\rm sat}$
is the saturated $Q$ value to which the disc settles in the presence of both the external irradiation and heating
through dissipating the gravitational instability. Using shearing sheet simulations, \citet{rice11} showed that this relationship
is indeed reasonable, and the results are illustrated in Figure \ref{fig:alpha_Qirr} (from \citealt{rice11}) which
shows how the effective $\alpha$ varies with increasing levels of external irradiation, when the underlying cooling timescale
is $\beta = 9$.  

The important point here is that if there are other sources of heating, the strength of the instability is not set by the cooling 
timescale, $\beta$, alone. In the presence of other heating sources, the strength of the instability, and the magnitude of $\alpha$,
will be smaller than if the only heating source is the gravitational instability. 

\subsection{Local vs non-local}
Another assumption underlying the idea that angular momentum transport via the graviational instability can be treated
as essentially viscous, is that the shear stress (or, equivalently, the viscous $\alpha$) depends primarily on the local conditions in the disc.
Gravity is, however, inherently global, and so such an assumption isn't strictly correct.  \citet{balbus99} have shown that the
energy equation for a self-gravitating disc has extra terms associated with global wave transport. They do, however, suggest that
self-gravitating discs that hover near marginal gravitational stability may behave in a manner that is well described by a 
local $\alpha$ model.

Early global simulations \citep{laughlin94} seemed to indicate that simple $\alpha$ models could reproduce the density evolution
of self-gravitating accretion discs.  Such simulations, however, initially ignored heating and cooling processes in the disc, which play
a fundamental role in setting the quasi-steady nature of such systems \citep{gammie01}.  Imposing a $\beta$-cooling formalism 
in global disc simulations, \citet{lodato04} illustrated that quasi-steady, self-gravitating discs do indeed behave like $\alpha$ discs
and that the local model is a reasonable approximation.

The local nature of self-gravitating angular momentum transport does, however, depend on the mass of the disc relative to the 
mass of the central object.  As the disc-to-star mass ratio increases, the disc becomes thicker and global effects are likely
to become more important.  \citet{lodato04} only considered discs that had masses less than one quarter the mass of the central star.
Later work \citep{lodato05} considered much higher mass ratios.  What was found was that as the disc-to-star mass ratio 
increases, lower-order spiral modes start to dominate.  For mass ratios of $\sim 0.5$, a transient $m = 2$ spiral develops that 
rapidly redistributes mass and then allows the disc settle into a quasi-steady, self-regulated state. For mass ratios approaching unity, however, the
disc never seems to settle into some kind of quasi-steady, self-regulated state.  However, it does appear to 
satisfy the basic criteria in a time-averaged sense.

The basic picture therefore appears to be that for mass ratios below $\sim 0.5$ a self-gravitating disc will rapidly settle into
a quasi-steady, self-regulated state in which $Q \sim 1$, heating and cooling are in balance, and in which angular momentum 
transport is pseudo-viscous and can be described by an effective $\alpha$. As the mass ratio increases, lower $m$-modes begin to dominate,
producing global spiral density waves that either rapidly redistibute mass, allowing the disc to then settle into a quasi-steady,
self-regulated state, or - for sufficiently high mass ratios - that have time varying amplitudes leading to the disc only being 
quasi-steady in a time averaged sense.

\subsection{Realistic cooling}
A great deal of work on the evolution of self-gravitating accretion discs has used the basic $\beta$-formalism when
implementing cooling.  This, however, is clearly not a realistic representation of how such discs will actually cool,
and was never intended to be so.  It is simply a basic way in which to represent cooling so as to investigate how such
discs evolve under a range of different cooling scenarios.

It's clear that in reality, a self-gravitating disc will not have a cooling function that can be described via a single
$\beta$ parameter \citep{pickett03} and that the actual evolution may depend on the form of the cooling function \citep{mejia05}. 
A number of studies have indeed used more realistic cooling formalisms. For example, \citet{boss01,boss03} and \citet{mayer07} 
used simulations with radiative transfer to understand the fragmentation of self-gravitating discs, \citet{cai06} considered how the disc
metallicity might influence its evolution, \citet{boley06} considered the quasi-steady evolution of discs with realistic cooling and
realistic opacities, and \citet{meru10} probed the parameter space where disc fragmentation may occur, using three-dimensional hydrodynamical
simulations with radiative transfer.

More recently \citet{forgan11a} also considered the evolution of self-gravitating discs using a more realistic cooling formalism, but
with the goal of trying to understand its quasi-steady evolution and, in particular, if the local approximation still applies when the 
cooling is more realistic. They performed three-dimensional Smoothed Particle Hydrodynamic (SPH) simulations of discs around young stars in which a radiative transfer formalism \citep{stamatellos07, forgan09} was used to estimate how
the disc cooled.  

Their results suggest that for discs with masses below about half that of the central star, the general behaviour is well described
by the local approximation.  An estimate of the local cooling time, when $Q \sim 1$, can be used to determine the local value of
$\alpha$.  However, as the mass ratio inceases, the behaviour becomes more global, and the local cooling time is not a good indicator
of the effective $\alpha$, except in a time-averaged sense.  

Given the above, it is possible to use semi-analytic calculations to determine the realistic structure of quasi-steady,
self-gravitating accretion discs \citep{rafikov09,clarke09,rice09}, at least for those with masses less than about half that of the
central star. For example, for a given opacity one can determine the cooling rate
for a given surface density and, hence, the local effective viscosity ($\alpha$).  Using standard one-dimensional accretion disc
models \citep{pringle81} one can then determine the surface density profile for a disc accreting at a specified accretion rate.  

\begin{figure}
\begin{center}
\includegraphics[width=8.5cm]{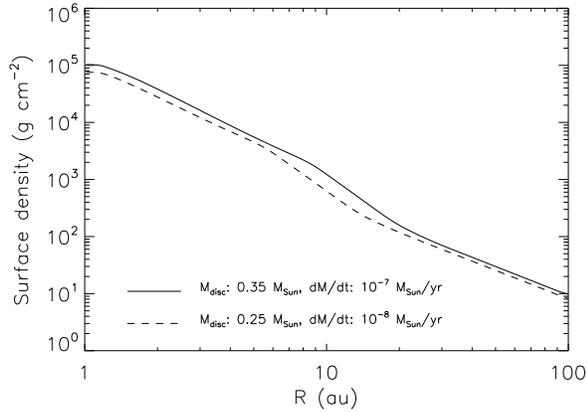}
\caption{Examples of two surface density profiles for self-gravitating discs that are in a quasi-steady state. In these examples,
there is no external irradiation and standard opacities are assumed.  For given opacities, the profile is essentially unique for
a specific mass accretion rate.  The mass accretion rates here are $\sim 10^{-7}$ M$_\odot$/year (solid line) and $\sim 10^{-8}$ M$_\odot$/year 
(dashed line) and the disc masses are $\sim 0.35$ M$_\odot$ (solid line) and $\sim 0.25$ M$_\odot$ (dashed line). This illustrates 
that a relatively small change in disc mass can produce a substantial change in accretion rate, indicating that disc self-gravity
is only likely to play an important role in angular momentum transport when these systems are very young, and the disc mass 
is relatively high.}
\label{fig:quasi-steady-sdens}
\end{center}
\end{figure}

For example, Figure \ref{fig:quasi-steady-sdens} shows the surface density profile of a quasi-steady, self-gravitating disc
around a 1 M$_\odot$ star accreting at $\sim 10^{-7}$ M$_\odot$/year (solid line) and accreting at $\sim 10^{-8}$ M$_\odot$/year (dashed line),
for standard opacities \citep{bell94} and in the absence of external irradiation.  Given the opacities, the profile is essentially
unique for a specific mass accretion rate. What Figure \ref{fig:quasi-steady-sdens} also shows is that there is quite a strong mass dependence. Assuming an outer disc
radius of 100 AU, the disc mass of the disc accreting at $\sim 10^{-7}$ M$_\odot$/year is $\sim 0.35$ M$_\odot$, while that of the disc
accreting at $\sim 10^{-8}$ M$_\odot$/year is $\sim 0.25$ M$_\odot$. Less than a factor of two change in disc mass, produces about an 
order of magnitude change in accretion rate.

This has a number of consequences.  To explain TTauri-like, and higher, accretion rates, self-gravitating discs need to be quite massive,
relative to the mass of the central star. Given that disc lifetimes are typically 5 Myr, or less \citep{haisch01}, self-gravity is unlikely
to explain accretion rates during the later stages of a discs lifetime, since it would imply much longer lifetimes than those observed.  
Similarly, observed spiral structures in protostellar discs are unlikely to be due to disc self-gravity unless the disc mass is still 
relatively high \citep{dong15}. For discs with masses less than $\sim 0.1$ M$_\odot$, the instability is likely to be weak, is unlikely 
to generate the stresses that can dominate the angular momentum transport process, and is unlikely to have spiral density waves that 
are sufficiently strong so as to be observable. 

\section{FRAGMENTATION}
The previous section focussed on the quasi-steady evolution of self-gravitating accretion discs.  For sufficiently unstable discs,
however, one possible outcome is that they fragment to form bound objects.  It has been suggested that these could be protoplanets
in discs around young stars \citep{boss98}, or protostars in discs around supermassive black holes \citep{levin03}. 

\begin{figure}
\begin{center}
\includegraphics[width=8cm]{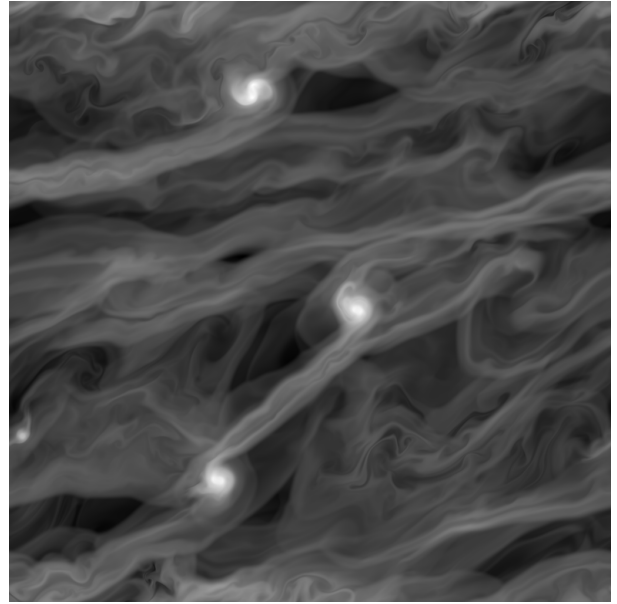}
\caption{A snapshot showing the surface density structure in a shearing sheet simulation with $\beta = 2$ and $\gamma = 2$. 
There are a number of dense clumps, indicating that the rapid cooling in this simulation has lead to the disc fragmenting,
rather than settling into a quasi-steady, self-regulated state.}
\label{fig:slice_2_1k}
\end{center}
\end{figure}

Again using shearing sheet simulations, \citet{gammie01} showed that a disc will fragment if the cooling time is sufficiently fast; 
in this case the fragmentation boundary being at $\beta = 3$. This is illustrated in Figure \ref{fig:slice_2_1k} which shows 
the surface density structure in a shearing sheet simulation with $\beta = 2$ and $\gamma = 2$. It is clear that there are a number
of dense clumps forming in the disc. 

A similar fragmentation boundary was obtained using three-dimensional, global simulations \citep{rice03}. However, as Equation
(\ref{eq:alpha_gammie}) shows, the effective $\alpha$ depends on the specific heat capacity $\gamma$. Consequently it was 
shown that the fragmentation boundary is actually set by a maximum $\alpha$ that can be sustained in a quasi-steady, self-regulated
disc, not by the cooling time specifically \citep{rice05}. For example, the cooling time for fragmentation if $\gamma = 2$ is smaller
than for $\gamma = 1.4$. 

Additionally, one has to be careful because all of this early work was done under the assumption that
dissipation of the gravitational instability was the only heating source.  As discussed above, the presence of external irradiation
weakens the instability and reduces the effective $\alpha$ \citep{rice11}. Therefore it is more reasonable to regard the fragmentation
boundary as being associated with the maximum self-gravitating $\alpha$ that a disc can sustain in a self-regulated, quasi-steady state, 
than as the minimum cooling time that can be imposed without the disc fragmenting. 

Furthermore, the above analyses have all used simulations in which the local approximation is reasonably valid.  As discussed
above, as the disc becomes more massive (relative to the mass of the central object) global modes start to dominate, and the
local approximations are only valid in a time averaged sense.  When global modes start to dominate, it appears that a self-gravitating
disc can sustain larger $\alpha$ values without fragmenting, than would be expected based on the local approximation only \citep{lodato05}. 
Simulations of collapsing cloud cores with radiative transfer also indicate that the behaviour is quite different to what would
be expected based on the local approximation \citep{forgan11a, tsukamoto15}.  Discs that don't fragment can have $\alpha$ values
considerably higher than the local approximation would suggest is possible. As \citet{forgan11a} illustrates, however,
the Reynolds stress, driven by velocity shear from material from the infalling envelope contacting the disc, dominates, and the
gravitational component of the stress still typically produces $\alpha$ values less than 0.1. 

\subsection{Perturbation amplitudes}
\begin{figure}
\begin{center}
\includegraphics[width=8.5cm]{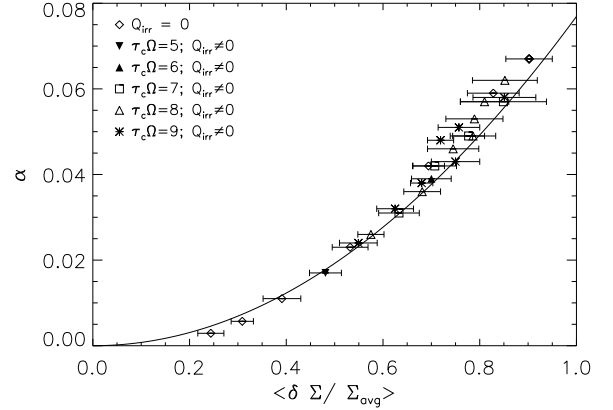}
\caption{A figure showing time-averaged $\alpha$ plotted against the mean perturbation amplitude $\left( \left< \delta \Sigma \right> / \left< \Sigma \right> \right)$, for a 
variety of different shearing sheet simulations.  It's clear that there is a relationship between $\alpha$ and the perturbation amplitudes,
with the perturbation amplitudes increasing with increasing $\alpha$. In the absence of external irradiation ($Q_{\rm irr} = 0$), this would also
indicate a relationship between the perturbation amplitudes and $\beta$ \citep{cossins09}, but in the presence of external irradiation
($Q_{\rm irr} \ne 0$) it is $\alpha$ that sets the perturbation amplitude, not simply $\beta$. (Figure from \citealt{rice11}.)}
\label{fig:alpha_sigrms}
\end{center}
\end{figure}

If we focus again on the case where the local approximation is valid, one way to understand fragmentation is to 
consider the relationship between $\alpha$ and the perturbation amplitudes. \citet{cossins09} show that the transport of
energy and angular momentum, by the spiral waves driven by the gravitational instability, depends on the surface density
perturbation $\delta \Sigma$. Since the angular momentum transport can be described via an effective $\alpha$, which itself
depends on the cooling timescale $\beta$, \citet{cossins09} show that the perturbation amplitudes depend on $\beta$ through
\begin{equation}
\frac{\left< \delta \Sigma \right>}{\left< \Sigma \right>} \simeq \frac{1.0}{\sqrt{\beta}}.
\label{eq:perturbations_beta}
\end{equation} 
Similarly, you can cast this as \citep{rice11}
\begin{equation}
\frac{\left< \delta \Sigma \right>}{\left< \Sigma \right>} \simeq \sqrt{\alpha}.
\label{eq:perturbations_alpha}
\end{equation}
Figure \ref{fig:alpha_sigrms} (taken from \citealt{rice11}) shows a plot of time-averaged $\alpha$ against $\left< \delta \Sigma \right> / \left< \Sigma \right>$ 
illustrating that as $\alpha$ increases the perturbation amplitudes increase. Figure \ref{fig:alpha_sigrms} also
shows results from simulations with no external irradiation ($Q_{\rm irr} = 0$) and with external irradiation 
($Q_{\rm irr} \ne 0$) illustrating that it is $\alpha$ that determines the perturbation amplitudes, not 
simply the cooling time $\beta$. 

Essentially, the basic picture is that as the self-gravitating $\alpha$ increases, the perturbations become larger and, 
there will be an $\alpha$ value beyond which the perturbations collapse and the disc fragments. Most simulations suggest that
this occurs at an $\alpha$ value of about 0.06 \citep{rice05}, but this should probably not be seen as some kind of hard boundary,
as we'll discuss in more detail later.

\subsection{Convergence}
Recently, it has been suggested that the fragmentation boundary does not converge to a fixed $\alpha$ value as the numerical
resolution of the simulation increases \citep{meru11, meru12, baehr15}. There have been a number of attempts to explain this. \citet{lodato11}
and \citet{meru12}  suggest that numerical viscosity may play a larger role than previously thought (e.g., \citealt{murray96}) and that, consequently, simulations will need to consider higher resolutions than have been used in the past.  \citet{paardekooper11} show that the lack
of convergence could be related to edge effects in simulations with very smooth initial conditions. This apparent non-convergence has, however,  
lead to a suggestion that fragmentation may occur for very long cooling times. \citet{meru12} do, however, extrapolate their results to suggest 
that they may be tending towards convergence and that the critical value may be around $\beta \sim 30$. 

One issue with the idea that fragmentation could occur for very long cooling times is, however, based on our basic understanding of the underlying processes.  As discussed above, in the absence
of other heating sources, there is a relationship between cooling time and $\alpha$, such that as the cooling time gets longer, 
the effective $\alpha$ value gets smaller.  However, this also implies that the perturbation amplitudes get smaller. 
So, one issue would be how do discs fragment if the perturbations are tiny?  

\begin{figure}
\begin{center}
\includegraphics[width=8.5cm]{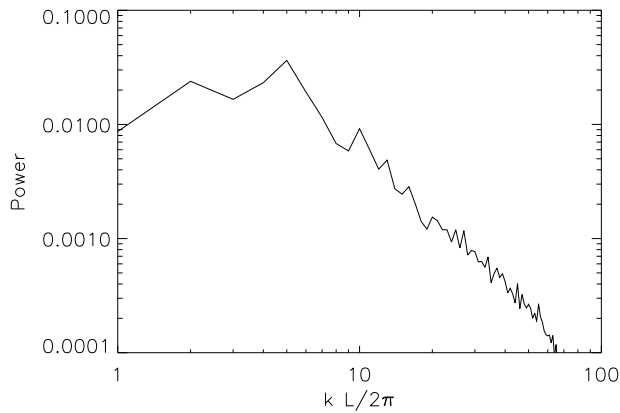}
\caption{Figure showing the power spectrum of the perturbations in a shearing sheet simulation with $\beta = 10$. 
The length of each side of the sheet is $L$, and so this figure shows that most of the power is at scales a few times smaller
than the sheet size and that there is very little power at very small scales.}
\label{fig:powerspec}
\end{center}
\end{figure}

Also, if fragmentation at longer cooling times
requires higher resolution, then it implies that this fragmentation is occuring on scales that are not resolved by the lower
resolution simulations.  However, the Jeans mass/radius at $Q \sim 1$ is typically resolved in at least part of the 
simulation domains for these lower resolution simulations \citep{rice14, tsukamoto15}.  This would suggest that at these 
longer cooling times, large amplitude, small-scale perturbations are driving $Q << 1$ and hence producing fragments that are unresolved
in the lower resolution simulations. In shearing sheet simulations, however, the power typically decreasing with
increasing wavenumber, $k$, as shown in Figure \ref{fig:powerspec}.  The issue then becomes why at long cooling times 
fragmentation happens at scales where there is much less power than at the scales at which fragmentation occurs when the
cooling time is short?    

Consequently, \citet{rice12}, suggest that the effect is largely numerical and is related to the manner in which cooling has typically
been implemented in SPH simulations.  More recently an SPH simulation with 10 million particles and with a modified form
of the standard $\beta$ cooling formalism, was shown not to fragment
for $\beta = 10$ \citep{rice14}, well below the values suggested by \citet{meru11} and \citet{meru12}. \citet{michael12} also considered convergence
in a three-dimensional grid-based code and, although they didn't test for the fragmentation boundary specifically, they did find that
simulations with $\alpha < 0.06$ typically did not fragment.  Hence it seems that it's important to be careful of how the cooling is 
implemented and to avoid implementations that may promote fragmentation.

There has been some attempt to try and see if there is a physical explanation for this lack of convergence. In particular, is
there evidence that simulations are suppressing fragmentation at small scales?  One possibility,
suggested by \citet{young15}, is that there are two modes
of fragmentation.  One mode operates at the free-fall time and requires rapid cooling. The other is a mode in which fragments form 
via quasi-static collapse and requires that these fragments then contract to a size where they can survive disruption by spiral shock waves. 
\citet{young15} suggest that many two-dimensional simulations smooth gravity on the Jeans scale, inhibiting contraction to smaller scales,
and hence preventing this second mode from operating. However, they also show that even if this second mode does operate,
it probably still requires relatively fast cooling ($\beta \lesssim 12$), but with quite a large uncertainty. 

Part of the motivation behind \citet{young15} was a suggestion that turbulent discs are never stable against fragmentation 
\citep{hopkins13} and, in particular, that rare high-amplitude, small-scale fluctuations could lead to gravitational collapse.  
This suggestion, however, is not entirely consistent with what we're discussing here.   The \citet{hopkins13}
analysis considered discs that were typically assumed to be isothermal, essentially implying that $\beta = 0$.  

Also, the turbulence they were considering was not the gravito-turbulence driven by the instability itself.  The standard analysis of
self-gravitating discs considers a scenario in which the quasi-turbulent structures are a self-consistent
response to the growth of the instability and depend - as discussed already - on the thermal balance
in the disc.  \citet{hopkins13} were considering a situation in which isothermal turbulence is present in a disc in which
self-gravity might also operate. In fact, the presence of such turbulence - if the disc were non-isothermal - should act as an
extra heating source, which would then be expected to reduce the magnitude of the gravitational instability, rather than enhance it 
\citep{rice11}.

\subsection{Triggered fragmentation}
Even though the analysis in \citet{hopkins13} is not quite consistent with the basic picture presented here, it is still
interesting from the perspective of whether or not fragmentation could be triggered. This could be via some other form
of turbulence (as suggested by \citealt{hopkins13}) or via some kind of external perturbation.  Some early work (e.g.,
\citealt{boffin98,watkins98}) suggested that stellar encounters could promote fragmentation.  Such early simulations typically,
however, assumed an isothermal equation of state, which is known to promote fragmentation \citep{pickett98,pickett00}.  
Later work suggested that the inclusion of compressive and shock heating \citep{nelson00,
lodato07} would tend to prohibit encounter-driven fragmentation.  This was further confirmed using simulations with 
a realistic radiative transfer formalism \citep{forgan09}.

The basic argument is that, typically, the encounter will compress the disc on a dynamical timescale,
which is much shorter than the thermal timescale.  Consequently, the encounter would be expected to heat -
and stablise - the disc, rather than produce fragmentation.  Fragmentation would only be expected to occur if   
the cooling time can be reduced by the compression \citep{whitworth07}. This may be possible if the temperature
change pushes the gas into a regime where the opacity is reduced (the opacity-gap). However, the corresponding
increase in gas density tends to act in the opposite sense, and so this may still make triggered fragmentation unlikely 
\citep{johnson03}.  There has, however, been a recent suggestion \citep{meru15} that fragmentation in the inner parts of dics 
may be triggered by fragments that form, initially, in the outer parts of discs, and \citet{thies10} suggest
that tidal encounters might trigger fragmentation in massive, extended protostellar discs.  Hence, there may be scenarios
under which fragmentation could be triggered.

\subsection{Stochasticity}
\begin{figure}
\begin{center}
\includegraphics[width=8.0cm]{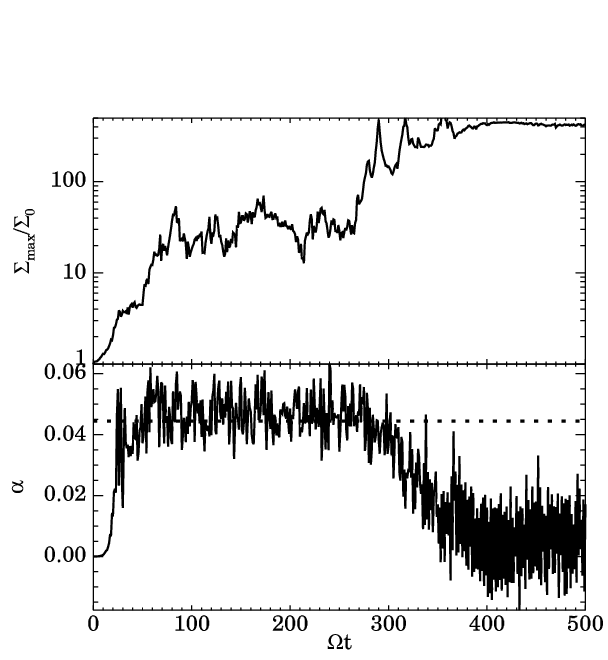}
\caption{Figure - from \citet{paardekooper12} - showing the maximum surface density relative to the initial surface density ($\Sigma_{\rm max}/ \Sigma_o$ - top panel) and $\alpha$ (bottom panel) against time, from a shearing sheet simulation that initially appears to have 
settled into a quasi-steady, self-regulated state.  However, at $\Omega t \sim 300$, $\alpha$ starts to decrease and the maximum surface density starts increasing, 
eventually reaching values many hundreds of times greater than it was initially.  The system, which appeared to have 
settled into a quasi-steady state, has now undergone fragmentation and this may be indicative of stochasticity.}
\label{fig:alpha_stoch}
\end{center}
\end{figure}

\citet{paardekooper12} have suggested that fragmentation may be stochastic. Their basic argument is that a quasi-steady
gravito-turbulent state consists of weak shocks and transient clumps that will contract on the cooling timescale ($\beta$). 
If such a clump can avoid being disrupted by one of the spiral shocks, then it could survive to become a bound fragment.
For example, Figure \ref{fig:alpha_stoch} - taken from \citet{paardekooper12} - shows the evolution of the maximum surface density ($\Sigma_{\rm max} / \Sigma_o$) and $\alpha$ 
in a shearing sheet simulation in which the disc appears to initially settle into
a quasi-steady, self-regulated state.  This is illustrated by both the maximum density and $\alpha$ settling to reasonably constant values, as indicated  
by the dashed line in the lower panel. 
However, at $\Omega t \sim 300$, $\alpha$ starts to decrease and the maximum density increases to to a value a few hundred times greater than the 
initial value, $\Sigma_o$.  This indicates that this system has undergone fragmentation after first appearing to settle to a quasi-steady state,
 and may be indicative of stochasticity.  

\citet{paardekooper12} find that
transient clumps are present in all of their simulations (up to $\beta = 50$) but that fragmentation only actually happens 
up to $\beta = 20$. \citet{hopkins13} suggest that, given sufficient time, stochastic fragmentation may actually occur both if the cooling time
is very long and the $Q$ value is very high.  However, as we suggest above, the scenario they're considering isn't quite
the same as that being discussed here. 

\citet{young16} argue that this stochasticity is consistent with the idea behind their suggestion that there are two modes of fragmentation \citep{young15}. 
In discs with rapid cooling, fragmentation can occur on the free-fall time.  In discs that are cooling slowly (for example, $\beta > 10 - 20$), fragmentation
can still occur if a clump can cool and contract prior to encountering a spiral shock. \citet{young16} find that the wait time
between shocks is not stongly dependent on the cooling time. Given that clumps contract on the cooling timescale, this means that
the likelihood of fragmentation decreases with increasing cooling time.  They also find an exponential decay in the 
likelihood of a patch remaining unshocked, and so the probability of a transient clump contracting sufficiently so as to survive
a shock encounter is effectively zero for very long cooling times.  Given this, even if fragmentation is stochastic, this is
unlikely to substantially change the conditions under which fragmentation actually occurs. Similarly, \citet{baehr15} impose a cooling
implementation that attempts to incorporate optical depth effects, and find that this also appears to inhibit stochastic fragmentation.

\subsection{Does fragmentation actually occur?}
\begin{figure}
\begin{center}
\includegraphics[width=8.5cm]{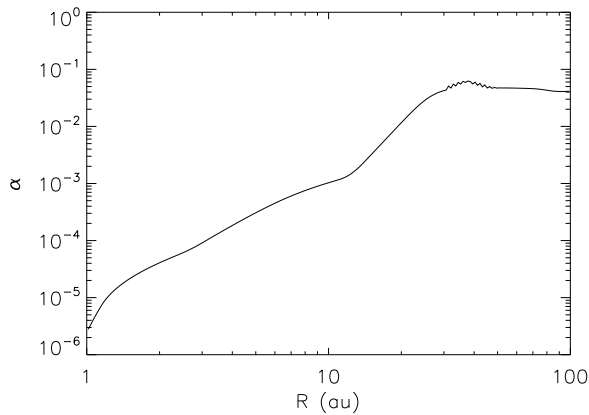}
\caption{Figure showing $\alpha$ against radius in a quasi-steady, self-gravitating disc with realistic cooling. The inner disc
is optically thick, leading to long cooling times and very small $\alpha$ values.  The $\alpha$ value increases with radius, 
reaching value where fragmentation might be possible in the outer parts of the disc.  Hence, we expect fragmentation to only be possible
in the outer parts of protostellar accretion discs.}
\label{fig:alpha_r}
\end{center}
\end{figure}

If we focus specifically on discs around young stars (protostellar discs) then we can determine 
where fragmentation could happen, and even if it does actually happen. It is likely 
that the inner regions of protostellar discs are optically thick and, consequently, 
have long cooling times.  This is illustrated in Figure \ref{fig:alpha_r} which shows $\alpha$ plotted against radius 
for a quasi-steady disc in which the cooling time at each radius is calculated using the actual optical depth, rather
than by simply assuming some $\beta$-value {\em a priori}. The $\alpha$ values are extremely small in the inner disc but increase
with increasing radius, reaching values where fragmentation may be possible in the outer parts of the disc. 
Consequently, we expect fragmentation to only occur in the outer parts of such systems \citep{matzner05, rafikov05, stamatellos08},
beyond $\sim 30 - 50$ AU.

Even if fragmentation is stochastic, or can occur for longer cooling times than we currently think, 
this has little impact when it comes to fragmentation in protostellar discs. The radial dependence of the
cooling time is very strong in the inner parts of these discs \citep{clarke09,rice09}.  Even a factor of
a few in the cooling time for fragmentation has little effect on where fragmentation likely happens \citep{young16}. 
If the opacity is significantly lower than interstellar, then the fragmentation radius could to move towards the inner parts
of the disc \citep{meru10}. However, it is still unlikely that fragmentation can take place inside $\sim 10 - 20$ AU 
in such discs \citep{meru10}

In addition to where we might expect fragmentation to occur, we can also estimate typical fragment masses \citep{boley10, forgan11b}.
Typically, the initial fragment masses are a few Jupiter masses.  Given that we expect fragmentation to only occur in the
outer parts of protostellar discs, and to have initial masses of at least a few Jupiter masses, it's been suggested this could
explain some of the directly imaged planets on wide orbits \citep{kratter10,nero09}. It's also been suggested that in some cases such
objects could rapidly spiral into the inner disc \citep{baruteau11,michael11, malik15} and could lose mass via tidal stripping \citep{nayakshin10,boley10}. 
This could potentially then be the origin of some of the known exoplanets, even terrestrial/rocky exoplanets if the cores can survive disruption 
\citep{nayakshin11}.  If not, the disruption of an embryo could lead to the formation of planetesimal belts \citep{nayakshin12} 
or change the overall disc chemistry \citep{boley10}. In some cases, however, it may be possible for the protoplanet to open a gap
and survive \citep{stamatellos15}. 

\citet{forgan13}, however, argue that typically fragments are either destroyed, or remain at large radii with relatively 
high masses (many Jupiter masses).  According to their analysis, the formation of planetesimal belts or terrestrial planets via this mechanism
is also rare.  The high destruction rate, however, is consistent with other simulations \citep{boleydurisen10, zhu12} and could lead 
to outbursts, such as the FU Orionis phenomenon \citep{vorobyov05, dunham12}. \citet{galvagni14}, however, suggest that a large 
fraction of the clumps could survive inward migration and that such a process could explain a reasonable fraction of the known
population of `hot' Jupiters and other gas giant planets. 

In addition to the rapid inspiral of fragments, it's also possible that objects forming via fragmentation at large radii
could be scattered by stellar, or other, companions \citep{forgan15}.  Some of these could then be circularised onto a
short-period orbit, via tidal interactions with the parent star \citep{lin96}, producing either a `hot' Jupiter or a 
proto-hot Jupiter \citep{dawson12}. 
\citet{rice15}, however, argue that the population of known `hot' and proto-hot Jupiters is inconsistent with this being common
and suggests that disc fragmentation rarely forms planetary-mass objects. This is also consistent with the relatively low frequency of known wide orbit
planetary and brown dwarf companions \citep{biller13,brandt14}.

\section{DISCUSSION AND CONCLUSION}
We've tried here to summarise our current understanding of the evolution of self-gravitating accretion discs, 
focusing - in particular - on those around youngs protostars.  When self-gravity is important, we
expect such discs to either settle into a quasi-steady, self-regulated state in which heating and cooling
are in balance and $Q \sim 1$ \citep{paczynski78}, or - if the instability is particularly strong - to fragment to form bound objects. 

Typically the boundary between a disc fragmenting, or settling into a quasi-steady state, is determined by the rate at
which it is losing energy, with rapid cooling potentially leading to fragmentation \citep{gammie01}.  The exact boundary does, however, 
depend on whether or not there is an additional heating source \citep{rice11}, and on the mass of the disc relative to the mass of the
central object; due to the global nature of the gravitational instability \citep{balbus99},  very massive discs may avoid fragmentation under conditions that may lead to fragmentation in less massive discs \citep{lodato05, forgan11a}.  Similarly, very massive discs may not settle into 
a true quasi-steady, self-regulated state, but instead show variability, and are only quasi-steady in a time averaged sense.
It is also possible that fragmentation itself is stochastic and could, at times,
occur for longer cooling times than this basic analysis suggests \citep{paardekooper12}.  

In protostellar discs, however, the exact fragmentation boundary is not that relevant when it comes to determining where fragmentation
might happen. The inner parts of such discs are so optically thick, and cool so slowly, that fragmentation is only actually likely in the outer
parts of these discs \citep{rafikov05}.  Even if fragmentation could occur for slightly longer cooling times than originally thought, or if
fragmentation is stochastic, this is unlikely to significantly influence where fragmentation can occur \citep{young16}. Although, reductions
in opacity can bring the fragmentation region closer to the central star \citep{meru10,rogers12}, it's still unlikely close
to the central star.  Essentially, it is very difficult for there to be conditions such that fragmentation could occur inside $\sim 20$ AU.

There are, however, still aspects that are uncertain. It might seem that the lack of convergence in self-gravitating
disc simulations \citep{meru11} is most likely numerical \citep{lodato11, meru12, rice12, rice14}, but this is still not fully resolved. 
Similarly, there is evidence for stochasticity \citep{paardekooper12} and a potential explanation for this \citep{young16}, but
there are still aspects of this that are uncertain. However, our basic understanding seems robust. Self-gravitating discs will
tend to settle into a quasi-steady, self-regulated state in which heating balances cooling, in which the transport of
angular momentum can be described as pseudo-viscous \citep{lodato04}, and in which the amplitude of the perturbations
depends on the magnitude of the effective viscosity $\alpha$ \citep{cossins09}. If the perturbations become
very large, then the disc may undergo fragmentation.

It's also unclear as to whether or not fragmentation ever actually occurs in protostellar discs. There are suggestions
\citep{kratter10, nero09} that it may explain some of the directly imaged exoplanets on wide-orbits \citep{marois08} and
suggestions that some may spiral inwards to smaller orbital radii than where they form \citep{nayakshin10}. On the other
hand, the paucity of directly imaged planets on wide-orbits \citep{biller13} and the properties of the known closer-in
exoplanets, suggests that disc fragmentation may rarely form planetary-mass objects \citep{rice15}. 

However, disc self-gravity may still play a crucial role in angular momentum transport during the earliest stages of star formation, and
the resulting density waves may then play a role in the growth of planet building material \citep{rice04, gibbons12}. So, even
if the gravitational instability does not play a significant role in forming planets directly, it may still play an important 
role in the growth of the building blocks crucial for planet formation.
 
\begin{acknowledgements}
The author would like to thank the organisers of the Disc Dynamics and Planet Formation meeting, held in Cyprus,
that partly motivated this paper.  The author would also like to thank the reviewer for a very thorough and detailed review
that helped to improve the paper, and would like to acknowledge useful comments from Phil Armitage.  KR 
gratefully acknowledges support from STFC grant ST/M001229/1.  Some of the research leading to these results
also received funding from the European Union Seventh Framework Programme (FP7/2007-2013) under grant agreement number
313014 (ETAEARTH). 
\end{acknowledgements}


\begin{thebibliography}{}
\bibitem[Baehr \& Klahr(2015)]{baehr15}
Baehr, H. \& Klahr, H., 2015, ApJ, 814, 155

\bibitem[Balbus \& Papaloizou(1999)]{balbus99}
Balbus, S.A. \& Papaloizoum, J.C.B., 1999, ApJ, 521, 650

\bibitem[Baruteau, Meru \& Paardekooper(2011)]{baruteau11}
Baruteau, C., Meru, F. \& Paardekooper, S.-J., 2011, MNRAS, 416, 1971

\bibitem[Bell \& Lin(1994)]{bell94}
Bell, K.R. \& Lin, D.N.C., 1994, ApJ, 427, 987

\bibitem[Biller et al.(2013)]{biller13}
Biller, B.A., et al., 2013, ApJ, 777, 160

\bibitem[Boffin et al.(1998)]{boffin98}
Boffin, H.M.J., Watkins, S.J., Bhattal, A.S., Francis, N. \& Whitworth, A.P., 1998, MNRAS, 300, 1189

\bibitem[Boley et al.(2010)]{boley10}
Boley, A.C., Hayfield, T., Mayer. L. \& Durisen, R.H., 2010, Icarus, 207, 509

\bibitem[Boley \& Durisen(2010)]{boleydurisen10}
Boley, A.C. \& Durisen, R.H., 2010, ApJ, 724, 618

\bibitem[Boley et al.(2006)]{boley06}
Boley, A.C., Mej\'{i}a, A.C., Durisen, R.H., Cai, K., Pickett, M.K. \& D'Alessio, P., 2006, ApJ, 651, 517

\bibitem[Bonnell \& Rice(2008)]{bonnell08}
Bonnell, I.A. \& Rice, W.K.M., 2008, Science, 321, 1060

\bibitem[Boss(2003)]{boss03}
Boss, A.P., 2003, ApJ, 599, 577

\bibitem[Boss(2001)]{boss01}
Boss, A.P., 2001, ApJ, 551, L167

\bibitem[Boss(2000)]{boss00}
Boss, A.P., 2000, ApJ, 536, L101

\bibitem[Boss(1998)]{boss98}
Boss, A.P., 1998, Nature, 393, 141

\bibitem[Brandt et al.(2014)]{brandt14}
Brandt, T.D., et al., 2014, ApJ, 794, 159

\bibitem[Cai et al.(2006)]{cai06}
Cai, K., Durisen, R.H., Michael, S., Boley, A.C., Mej\'{i}a, A.C., Pickett, M.K. \& D'Alessio, P., 2006, ApJ, 636, L149

\bibitem[Clarke(2009)]{clarke09}
Clarke, C.J., 2009, MNRAS, 396, 1066

\bibitem[Cossins, Lodato \& Clarke(2009)]{cossins09}
Cossins, P., Lodato, G. \& Clarke, C.J., 2009, MNRAS, 393, 1157

\bibitem[Dawson \& Johnson(2012)]{dawson12}
Dawson, R.I. \& Johnson, J.A., 2012, ApJ, 756, 122

\bibitem[Dong et al.(2015)]{dong15}
Dong, R., Hall, C., Rice, K. \& Chiang, E., 2015, ApJ, 812, L32

\bibitem[Dunham \& Vorobyov(2012)]{dunham12}
Dunham, M.M. \& Vorobyov, E.I., 2012, ApJ, 747, 52

\bibitem[Forgan, Parker \& Rice(2015)]{forgan15}
Forgan, D.H., Parker, R.J. \& Rice, K., 2015, MNRAS, 447, 836

\bibitem[Forgan \& Rice(2013)]{forgan13}
Forgan, D.H. \& Rice, K., 2013, MNRAS, 432, 3168

\bibitem[Forgan et al.(2011)]{forgan11a}
Forgan, D., Rice, K., Cossins, P. \& Lodato, G., 2011, MNRAS, 410, 994

\bibitem[Forgan \& Rice(2011)]{forgan11b}
Forgan, D., \& Rice, K., 2011, MNRAS, 417, 1928

\bibitem[Forgan et al.(2009)]{forgan09}
Forgan, D., Rice, K., Stamatellos, D. \& Whitworth, A., 2009, MNRAS, 394, 882

\bibitem[Gammie(2001)]{gammie01}
Gammie, C.F., 2001, ApJ, 553, 174

\bibitem[Galvagni \& Mayer(2014)]{galvagni14}
Galvagni, M. \& Mayer, L., 2014, MNRAS, 437, 2909

\bibitem[Gibbons, Rice \& Mamatsashvili(2012)]{gibbons12}
Gibbons, P.G., Rice, W.K.M. \& Mamatsashvili, G.R., 2012, MNRAS, 426, 1444

\bibitem[Haisch, Lada \& Lada(2001)]{haisch01}
Haisch, K.E., Lada, E.A. \& Lada, C.J., 2001, ApJ, 553, L153

\bibitem[Hopkins \& Christiansen(2013)]{hopkins13}
Hopkins, P.F. \& Christiansen, J.L., 2013, ApJ, 776, 48

\bibitem[Johnson \& Gammie(2003)]{johnson03}
Johnson, B.M. \& Gammie, C.F., 2003, ApJ, 597, 131

\bibitem[Kratter, Murray-Clay \& Youdin(2010)]{kratter10}
Kratter, K.M., Murray-Clay, R.A. \& Youdin, A.N, 2010, ApJ, 710, 1375

\bibitem[Laughlin \& Bodenheimer(1994)]{laughlin94}
Laughlin, G. \& Bodenheimer, P., 1994, ApJ, 436, 335

\bibitem[Laughlin \& Rozyczka(1996)]{laughlin96}
Laughlin, G. \& Rozcyczka, M., 1996, apJ, 456, 279

\bibitem[Levin \& Beloborodov(2003)]{levin03}
Levin, Y. \& Beloborodov, A.M., 2003, ApJ, 590, L33

\bibitem[Lin, Bodenheimer \& Richardson(1996)]{lin96}
Lin, D.N.C., Bodenheimer, P. \& Richardson, D.C., 1996, Nature, 380, 606

\bibitem[Lin \& Pringle(1987)]{lin87}
Lin, D.N.C. \& Pringle, J.E., 1987, MNRAS, 225, 607

\bibitem[Lodato \& Clarke(2011)]{lodato11}
Lodato, G. \& Clarke, C.J., 2011, MNRAS, 413, 2735

\bibitem[Lodato et al.(2007)]{lodato07}
Lodato, G., Meru, F., Clarke, C.J. \& Rice, W.K.M., 2007, MNRAS, 374, 590

\bibitem[Lodato \& Rice(2004)]{lodato04}
Lodato, G. \& Rice, W.K.M., 2004, MNRAS, 351, 630

\bibitem[Lodato \& Rice(2005)]{lodato05}
Lodato, G. \& Rice, W.K.M., 2005, MNRAS, 358, 1489

\bibitem[Malik et al.(2015)]{malik15}
Malik, M., Meru, M., Mayer, L. \& Meyer, M., 2015, ApJ, 802, 56

\bibitem[Marois et al.(2008)]{marois08}
Marois, C., Macintosh, B., Barman, T., Zuckerman, B., Song, I., Patience, J., Lafreni\'ere, D. \& Doyon, R., 2008, Science, 322, 1348

\bibitem[Matzner \& Levin(2005)]{matzner05}
Matzner, C.D. \& Levin, Y., 2005, ApJ, 628, 817

\bibitem[Mayer et al.(2007)]{mayer07}
Mayer, L., Lufkin, G., Quinn, T. \& Wadsley, J., 2007, ApJ, 661, L77

\bibitem[Mej\'{i}a et al.(2005)]{mejia05}
Mej\'{i}a, A.C., Durisen, R.H., Pickett, M.K., \& Cai, K., 2005, ApJ, 619, 1098

\bibitem[Meru(2015)]{meru15}
Meru, F., 2015, MNRAS, 454, 2529

\bibitem[Meru \& Bate(2012)]{meru12}
Meru, F. \& Bate, M.R., 2012, MNRAS, 427, 2022

\bibitem[Meru \& Bate(2011)]{meru11}
Meru, F. \& Bate, M.R., 2011, MNRAS, 411, L1

\bibitem[Meru \& Bate(2010)]{meru10}
Meru, F. \& Bate, M.R., 2010, MNRAS, 406, 2279

\bibitem[Michael et al.(2012)]{michael12}
Michael, S., Steiman-Cameron, T.Y., Durisen, R.H. \& Boley, A.C., 2012, ApJ, 746, 98

\bibitem[Michael, Durisen \& Boley(2011)]{michael11}
Michael, S., Durisen, R.H. \& Boley, A.C., 2011, ApJ, 737, L42

\bibitem[Murray(1996)]{murray96}
Murray, J.R., 1996, MNRAS, 279, 402

\bibitem[Nayakshin \& Cha(2012)]{nayakshin12}
Nayakshin, S. \& Cha, S.-H., 2012, MNRAS, 423, 2104

\bibitem[Nayakshin(2011)]{nayakshin11}
Nayakshin, S., 2011, MNRAS, 410, L1

\bibitem[Nayakshin(2010)]{nayakshin10}
Nayakshin, S., 2010, MNRAS, 408, L36

\bibitem[Nayakshin, Cuadra \& Springel(2007)]{nayakshin07}
Nayakshin, S., Cuadra, J. \& Springel, V., 2007, MNRAS, 379, 21

\bibitem[Nelson, Benz \& Ruzmaikina(2000)]{nelson00}
Nelson, A.F., Benz, W. \& Ruzmaikina, T.V., 2000, 529, 357

\bibitem[Nero \& Bjorkman(2009)]{nero09}
Nero, D. \& Bjorkman, J.E., 2009, ApJ, 702, L163

\bibitem[Paardekooper(2012)]{paardekooper12}
Paardekooper, S.-J., 2012, MNRAS, 421, 3286

\bibitem[Paardekooper, Baruteau \& Meru(2011)]{paardekooper11}
Paardekooper, S.-J., Baruteau, C. \& Meru, F., 2011, MNRAS, 416, L65

\bibitem[Paczynski(1978)]{paczynski78}
Paczynski, B., 1978, Acta Astron., 28, 91

\bibitem[Pickett et al.(1998)]{pickett98}
Pickett, B.K., Benz, W., Adams, F.C. \& Arnett, D., 1998, ApJ, 502, 342

\bibitem[Pickett et al.(2000)]{pickett00}
Pickett, B.K., Cassen, P., Durisen, R.H. \& Link, R., 2000, ApJ, 529, 1034

\bibitem[Pickett et al.(2003)]{pickett03}
Pickett, B.K., Mej\'{i}a, A.C., Durisen, R.H., Cassen, P.M., Berry, D.K. \& Link, R.P., 2003, ApJ, 590, 1060

\bibitem[Pringle(1981)]{pringle81}
Pringle, J.E., 1981, ARA\&A, 19, 137

\bibitem[Rafikov(2009)]{rafikov09}
Rafikov, R.R., 2009, ApJ, 704, 281

\bibitem[Rafikov(2005)]{rafikov05}
Rafikov, R.R., 2005, ApJ, 621, 69

\bibitem[Rice et al.(2015)]{rice15}
Rice, K., Lopez, E., Forgan, D. \& Biller, B., 2015, MNRAS, 454, 1940

\bibitem[Rice et al.(2014)]{rice14}
Rice, W.K.M., Paardekooper, S.-J., Forgan, D.H. \& Armitage, P.J., 2014, MNRAS, 438, 1593

\bibitem[Rice, Forgan \& Armitage(2012)]{rice12}
Rice, W.K.M., Forgan, D.H. \& Armitage, P.J., 2012, MNRAS, 420, 1640

\bibitem[Rice et al.(2011)]{rice11}
Rice, W.K.M., Armitage, P.J., Mamatsashvili, G.R., Lodato, G. \& Clarke, C.J., 2011, MNRAS, 418, 1356

\bibitem[Rice, Mayo \& Armitage(2010)]{rice10}
Rice, W.K.M., Mayo, J.H. \& Armitage, P.J., 2010, MNRAS, 402, 1740

\bibitem[Rice \& Armitage(2009)]{rice09}
Rice, W.K.M. \& Armitage, P.J., 2009, MNRAS, 396, 2228

\bibitem[Rice, Lodato \& Armitage(2005)]{rice05}
Rice, W.K.M., Lodato, G. \& Armitage, P.J., 2005, MNRAS, 364, 56

\bibitem[Rice et al.(2004)]{rice04}
Rice, W.K.M., Lodato, G., Pringle, J.E., Armitage, P.J. \& Bonnell, I.A., 2004, MNRAS, 355, 543

\bibitem[Rice et al.(2003)]{rice03}
Rice, W.K.M., Armitage, P.J., Bate, M.R. \& Bonnell, I.A., 2003, MNRAS, 339, 1025

\bibitem[Rogers \& Wadsley(2012)]{rogers12}
Rogers, P.D. \& Wadsley J., 2012, MNRAS, 423, 1896

\bibitem[Safronov(1960)]{safronov60}
Safronov, V.S., 1960, Ann. Astrophys., 23, 979

\bibitem[Shakura \& Sunyaev(1973)]{shakura73}
Shakura, N.I. \& Sunyaev, R.A., 1973, A\&A, 24, 337

\bibitem[Stamatellos(2015)]{stamatellos15}
Stamatellos, D., 2015, ApJ, 810, L11

\bibitem[Stamatellos \& Whitworth(2008)]{stamatellos08}
Stamatellos, D. \& Whitworth, A.P., 2008, MNRAS, 392, 413

\bibitem[Stamatellos et al.(2007)]{stamatellos07}
Stamatellos, D., Whitworth, A.P., Bisbas, T. \& Goodwin, S., 2007, A\&A, 475, 37 

\bibitem[Thies et al.(2010)]{thies10}
Thies, I., Kroupa, P., Goodwin, S.P., Stamatellos, D. \& Whitworth, A.P., 2010, ApJ, 717, 577

\bibitem[Toomre(1964)]{toomre64}
Toomre, A., 1964, ApJ, 139, 1217

\bibitem[Tsukamoto et al.(2015)]{tsukamoto15}
Tsukamoto, Y., Takahashi, S.Z., Machida, M.N. \& Inutsuka, S., 2015, MNRAS, 446, 1175

\bibitem[Vorobyov \& Basu(2005)]{vorobyov05}
Vorobyov, E.I. \& Basu, S., 2005, ApJ, 633, L137

\bibitem[Watkins et al.(1998)]{watkins98}
Watkins, S.J., Bhattal, A.S., Boffin, H.M.J., Francis, N. \& Whitworth, A.P., 1998, MNRAS, 300, 1205

\bibitem[Whitworth et al.(2007)]{whitworth07}
Whitworth, A.P., Bate, M.R., Nordlund, {\AA}, Reipurth, B. \& Zinnecker, H., 2007, in Reipurth B., 
Keil K., eds, Protostars and Planets V. Univ. Arizona Press, Tucson, p. 459

\bibitem[Young \& Clarke(2016)]{young16}
Young, M.D. \& Clarke, C.J., 2016, MNRAS, 455, 1438 

\bibitem[Young \& Clarke(2015)]{young15}
Young, M.D. \& Clarke, C.J., 2015, MNRAS, 451, 3987

\bibitem[Zhu et al.(2012)]{zhu12}
Zhu, Z., Hartmann, L., Nelson, R.P. \& Gammie, C.F., 2012, ApJ, 746, 110

\end{thebibliography}

\end{document}